\documentclass[journal]{IEEEtran}

\usepackage{xcolor,soul,framed} 

\colorlet{shadecolor}{yellow}
\usepackage[pdftex]{graphicx}
\graphicspath{{../pdf/}{../jpeg/}}
\DeclareGraphicsExtensions{.pdf,.jpeg,.png}

\usepackage[cmex10]{amsmath}

\usepackage{array}
\usepackage{mdwmath}
\usepackage{mdwtab}
\usepackage{eqparbox}
\usepackage{url}
\usepackage{multicol}
\usepackage{float}
\usepackage{booktabs}
\usepackage{rotating}
\usepackage[style=ieee]{biblatex}
\usepackage{subcaption}
\usepackage[compatibility=false]{caption}
\usepackage{xcolor}
\usepackage{svg}

\usepackage{hyperref}
\hypersetup{colorlinks=true,linkcolor=blue,urlcolor=blue}
\colorlet{promptboxcolor}{gray!10}
\newenvironment{promptbox}{%
  \MakeFramed {\advance\hsize-\width \FrameRestore}%
}{\endMakeFramed}

\usepackage{placeins}

\begin{document}

\bstctlcite{IEEEexample:BSTcontrol}
    \title{Leveraging Large Language Models to Obscure Code Stylometry: A Comparative Study of GPT-3.5 and GPT-4}
    
\author{
\begin{tabular}{@{}c@{\hspace{1em}}c@{}}
\centering
Saman Pordanesh & Dr. Benjamin Tan \\
saman.pordanesh@ucalgary.ca & benjamin.tan1@ucalgary.ca
\end{tabular}
}

\markboth{UNIVERSITY OF CALGARY, SCHULICH SCHOOL OF ENGINEERING, UNDERGRADUATE RESEARCH THESIS, Winter 2024
}{S. Pordanesh}

\maketitle


\begin{abstract}

In the rapidly evolving field of software development, code stylometry
[1]—analyzing unique stylistic signatures of programmers—plays a crit-
ical role in authorship attribution and cybersecurity. Recent advancements
in artificial intelligence, particularly Large Language Models (LLMs) like
GPT-3.5 [2] and GPT-4 [3], have introduced new dimensions to this field,
challenging traditional stylometry techniques. This study investigates the
effectiveness of LLMs in altering code stylometry while preserving func-
tionality and evaluates the impact of various prompt engineering strategies.
Through comprehensive experiments, we assess how well these models
can obscure stylistic signatures to avoid detection by a Random Forest [4]
classifier trained for authorship attribution. The results reveal significant
differences in effectiveness between single-shot and multi-shot methods
and highlight the importance of detailed, structured prompts. Additionally,
functionality preservation checks demonstrate the challenges in maintain-
ing code integrity post-modification. This research provides critical insights
into the robustness of authorship attribution techniques against advanced
AI capabilities, informing future cybersecurity and software engineering
developments.

All codes and experiemnt results available at \url{https://github.com/sinapordanesh/LLMs_on_Code_Stylometry}

\end{abstract}

\begin{IEEEkeywords}
Large Language Models (LLMs), GPT-4, Code Detection, Code Classification, 
Statistical Learning, Random Forest, Code Stylometry, Author Detection
\end{IEEEkeywords}


\IEEEpeerreviewmaketitle

\section{Introduction}

\IEEEPARstart{I}{n} the rapidly evolving field of software development, the unique stylistic signatures of programmers, known as code stylometry \cite{caliskan1}, play a critical role in authorship attribution and cybersecurity. Code stylometry involves analyzing various features of source code to identify the authorship or any other identity feature like the origin country of the Code, which is particularly significant in tracing the origins of malicious software \cite{caliskan1}. Recent advancements in artificial intelligence, particularly Large Language Models (LLMs), have introduced new dimensions to this field, enabling the generation and transformation of Code in ways that challenge traditional stylometry techniques \cite{bukhari}.

This study is motivated by the need to understand how these LLMs, specifically GPT-3.5 \cite{gpt-3.5} and GPT-4 \cite{gpt4}, can be leveraged to obscure the stylistic signatures of Code while maintaining its functionality. Such capabilities have profound implications, potentially aiding malware authors in evading detection and complicating the task of cybersecurity professionals \cite{caliskan1, wang}.

The primary objectives of this research are to investigate the effectiveness of LLMs in altering code stylometry, assess the functionality preservation of modified Code, and evaluate the impact of different prompt engineering strategies \cite{afroz}. Through this study, we aim to fill existing gaps in the literature and provide insights into the robustness of authorship attribution techniques in the face of advanced AI capabilities \cite{brennan}.

Several vital questions guide our research: How effectively can LLMs alter code stylometry to evade detection? What are the differences in effectiveness between single-shot and multi-shot methods? How do different prompt engineering strategies influence the alteration of code stylometry?

The significance of this study lies in its contributions to cybersecurity and software engineering. By understanding the capabilities and limitations of LLMs in this context, we can inform the development of more robust authorship attribution techniques and contribute to the broader discourse on software security \cite{alsulami}. This thesis is structured as follows: the Literature Review summarizes existing research and key concepts; the Methodology details our dataset, prompt features, models, and experimental design; the Results section presents and analyzes our findings; the Discussion interprets the results, explores limitations, and outlines future works; and the Conclusion summarizes our findings and suggests avenues for future research.

\section{Background \& Literature Survey}

\subsection{Code Stylometry}
Code stylometry involves analyzing the stylistic fingerprints of programmers in their source code. These fingerprints can be used to attribute authorship, identify plagiarism, and even de-anonymize programmers. The foundational work in this field was significantly advanced by Caliskan et al. \cite{caliskan1, caliskan2}, who demonstrated that machine learning methods could be employed to de-anonymize programmers with high accuracy by using features derived from abstract syntax trees (ASTs) \cite{ast}. This approach proved robust against simple confusion techniques and scaled well to large datasets. Similarly, Caliskan et al. \cite{caliskan2} explored the persistence of coding styles through the compilation process, showing that executable binaries retain enough stylistic information to identify their authors. This highlights the depth at which coding styles are ingrained in a programmer's output, surviving even significant transformations like compilation.

Further research by Brennan and Greenstadt demonstrated the vulnerabilities of stylometric techniques to adversarial attacks \cite{brennan}. They found that both obfuscation (hiding one's style) and imitation (mimicking another's style) could drastically reduce the effectiveness of authorship recognition methods. This underscores the need for robust stylometric techniques to withstand intentional attempts to disguise coding styles.

\subsection{Authorship Attribution Techniques}
Various techniques have been developed to attribute authorship based on code stylometry. Traditional methods rely heavily on lexical and format features, such as keyword usage, indentation styles, and comment patterns. However, these methods are easily circumvented through simple obfuscation. Alsulami et al. introduced deep learning models, particularly Long Short-Term Memory (LSTM) \cite{lstm} networks, to automatically extract relevant features from ASTs \cite{alsulami}. This method eliminates the need for hand-crafted features and proves resilient to obfuscation, as it focuses on the structural syntactic features of the Code.

In practical applications, Afroz et al. applied stylometric analysis to identify users in underground forums, demonstrating high accuracy in detecting "doppelgängers" or users with multiple accounts \cite{afroz}. Their work highlights the potential of stylometry in cybersecurity, particularly in monitoring and understanding cybercriminal activities.

Bukhari et al. investigated the differentiation between AI-generated and human-generated code, showing that certain stylistic features could reliably distinguish between the two \cite{bukhari}. This is particularly relevant given the increasing use of AI in code generation, which poses new challenges for authorship attribution.

\subsection{Large Language Models and Code Generation}
The advent of large language models (LLMs) such as GPT-3.5 \cite{gpt-3.5} and GPT-4 \cite{gpt4} has introduced new dimensions to code generation and transformation. These models can generate code snippets that maintain functionality while altering stylistic features, potentially complicating authorship attribution. The study by Wang et al. on robust learning against relational adversaries explores how these models can be used to obfuscate code styles effectively \cite{wang}. Their framework, Normalize-and-Predict (N\&P), leverages input normalization to achieve robustness against adversarial transformations, highlighting the potential for LLMs to aid in evading authorship detection.

Furthermore, Caliskan et al. demonstrated that coding styles persist through compilation, suggesting that even compiled binaries can be de-anonymized using advanced stylometric techniques \cite{caliskan2}. This persistence poses a significant challenge for anonymizing Code, especially when LLMs generate or modify the Code.

%

\section{Methodology \& Experiment Design}
\label{sec:methodology}

\begin{figure}[H]
    \centering
    \includegraphics[width=1.0\columnwidth]{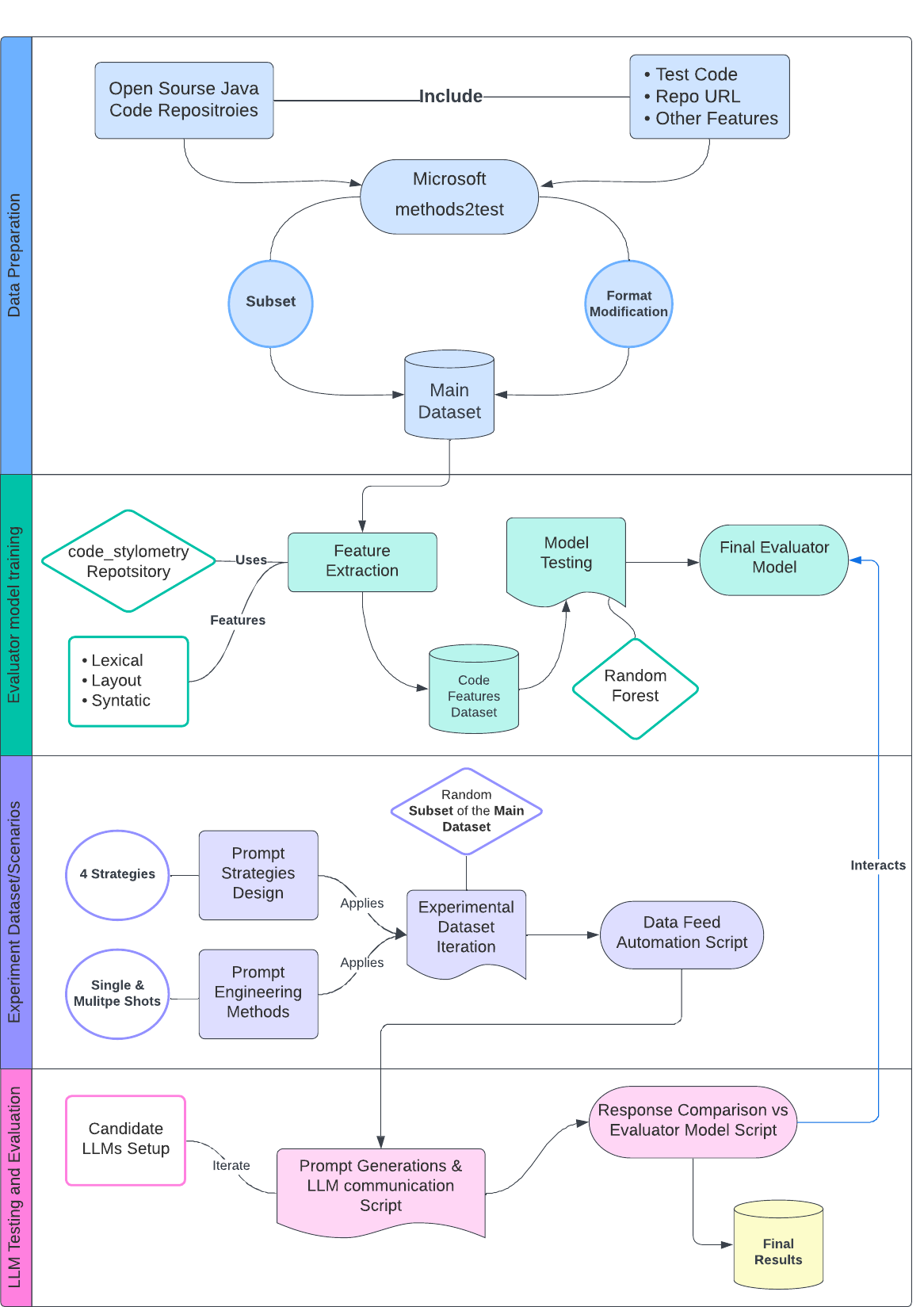} 
    \caption{Overview of the experimental design process.}
    \label{fig:experiment_design1}
\end{figure}

This research aims to investigate the effectiveness of Large Language Models (LLMs), specifically GPT-3.5 \cite{gpt-3.5} and GPT-4 \cite{gpt4}, in altering code stylometry to evade detection while maintaining the original functionality of the Code. This section outlines the comprehensive methodology and experimental design employed to achieve this goal. The process encompasses several critical phases, including dataset preparation, feature extraction, evaluator model training, and applying various prompt strategies to the LLMs. The overarching structure and flow of the experiment are illustrated in Figure~\ref{fig:experiment_design1}, providing a visual representation of the intricate steps involved.

\vskip -2em 

\vskip -8em 

\subsection{Datasets}
\begin{figure}[H]
    \centering
    \includegraphics[width=1\columnwidth]{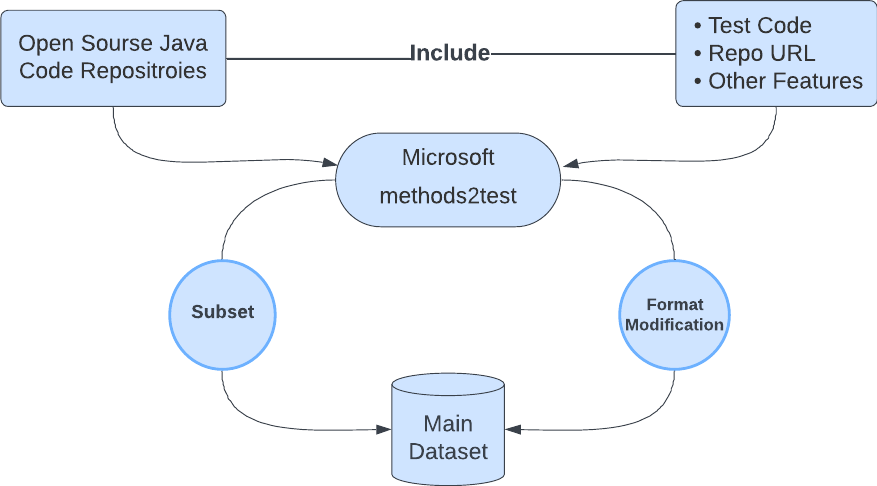}
    \caption{Overview of the data gathering process and dataset formation.}
    \label{fig:experiment_design}
\end{figure}

The dataset plays a pivotal role in this research project, serving as the foundation for the entire study. In the context of code stylometry and authorship attribution, the dataset's quality, comprehensiveness, and relevance directly influence the findings' accuracy and validity. The Methods2Test \cite{methods2test} dataset, specifically, was chosen for its extensive collection of Java code snippets and corresponding unit test cases, which are essential for training and evaluating our code stylometry model. Figure~\ref{fig:experiment_design} illustrates the data gathering process.

\subsubsection{Description of the Methods2Test Dataset}
The Methods2Test \cite{methods2test} dataset is a comprehensive collection of Java code snippets and their corresponding unit test cases, specifically designed to facilitate research in automated unit test generation and code stylometry. Tufano et al. introduced This dataset in the paper titled "Unit Test Case Generation with Transformers and Focal Context" \cite{tufano2020unit}. It represents the most extensive publicly available parallel corpus of test cases mapped to the corresponding focal methods. It comprises \textbf{780,944} unique test case pairs mined from \textbf{9,410} open-source Java projects hosted on GitHub.

The dataset collection process involved the following steps, based on what authors documented on the dataset repository \cite{methods2test} :

\begin{enumerate}
  \item \textbf{Source Selection}: The authors selected \textbf{9,410} public GitHub Java repositories that declared an open-source license, had been updated within the last five years, and were not forks.
  \item \textbf{Parsing and Metadata Extraction}: Each project was parsed using the tree-sitter parser, which collected metadata associated with the classes and methods identified within the project. This included method and class names, signatures, bodies, annotations, and variables.
  \item \textbf{Test Case Identification}: Classes containing at least one method with the Test annotation were marked as test classes.
  \item \textbf{Focal Method Mapping}: The focal class (the class under test) was identified for each test class. Heuristics such as path matching (a mirrored folder structure for Code and tests) and name matching were employed to map each test case to its related focal method.
  \item \textbf{Deduplication and Splitting}: Duplicate entries were removed, resulting in \textbf{780,944} unique test case pairs. The dataset was then split into training (80\%, 624,022 pairs), validation (10\%, 78,534 pairs), and test (10\%, 78,388 pairs) sets, ensuring no data leakage by keeping data points from the same repository within the same set.
\end{enumerate}

All dataset tuples were gathered in JSON format. For a better understanding of the Methods2Test dataset JSON structure, find more information in Appendix A (Figure. \ref{fig:JSON_structure}).

\subsubsection{Selection of Code Samples \& Format Modifications}
For this thesis, the Methods2Test dataset was utilized to train a code stylometry model (or evaluator model). Due to computational constraints and to maintain the project’s scope, only the test dataset (10\% of the total dataset - 78,388 pairs) was used. This selection was guided by the need to balance dataset size and the feasibility of running extensive experiments within the available resources.

We also needed to modify the original dataset format to better structure it for future machine-learning purposes. The criteria for sample modifications were as follows:

\begin{enumerate}
  \item \textbf{JSON to CSV Format}: The original subset of the dataset was created in the JSON format for each dataset sample. For our experimental implementation, we decided to use this dataset in the format of a CSV table. As a result, we transferred all of our datasets from JSON to CSV.
  \item \textbf{Initial Cleaning}: We initially deleted all rows with missing values like repo\_id, repository URL, Focal Method or Test Class codes after converting JSON to CSV file.
  \item \textbf{Assembled Code Collection}: As the original JSON file structure did not have Focal Method and Test Class codes assembled in a single column, we downloaded all codes from their original repositories using the provided URL and directory address for each tuple. We then added two new columns to our dataset: \texttt{focal\_class\_code} and \texttt{test\_class\_code}.
\end{enumerate}

\subsection{Stylometric Feature Extraction}
\begin{figure}[H]
    \centering
    \includegraphics[width=1\columnwidth]{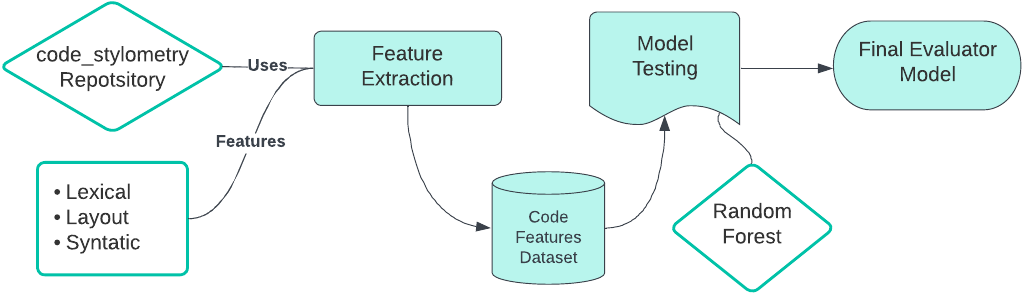}
    \caption{Overview of the feature extraction process.}
    \label{fig:experiment_design2}
\end{figure}

The process of extracting stylometric features from Code involves analyzing various characteristics of the source code that reflect a programmer's unique style. These features can be broadly categorized into lexical, layout, and syntactic features. The extraction process, illustrated in Figure~\ref{fig:experiment_design2}, was inspired by the methodology outlined in the paper "De-anonymizing Programmers via Code Stylometry" by Caliskan-Islam et al. \cite{caliskan1}. For this experiment, we utilized a Python repository \cite{rebryk} that implements these feature extractors.

\subsubsection{Lexical Features}
Lexical features pertain to the textual elements of the Code, including keywords, tokens, and literals. These features capture the frequency and usage patterns of specific words and symbols. The extracted lexical features are listed in Table~\ref{tab:lexical_features}.

\begin{table}[!t]
    \centering
    \caption{Lexical Features}
    \label{tab:lexical_features}
    \begin{tabular}{ll}
        \toprule
        \textbf{Feature} & \textbf{Explanation} \\
        \midrule
        NumKeyword & Count of specific keywords \\
        NumTokens & Total number of tokens \\
        NumLiterals & Count of literal values \\
        NumFunctions & Number of function definitions \\
        NumTernary & Number of ternary operators \\
        AvgLineLength & Average length of lines \\
        StdDevLineLength & Standard deviation of line lengths \\
        AvgParams & Average number of parameters per function \\
        StdDevNumParams & Standard deviation of parameters \\
        \bottomrule
    \end{tabular}
\end{table}

\subsubsection{Layout Features}
Layout features focus on the formatting and structural aspects of the Code, such as indentation, spacing, and line breaks. These features provide insight into the programmer's formatting habits. The extracted layout features are listed in Table~\ref{tab:layout_features}.

\begin{table}[!t]
    \centering
    \caption{Layout Features}
    \label{tab:layout_features}
    \begin{tabular}{ll} 
        \toprule
        \textbf{Feature} & \textbf{Explanation} \\
        \midrule
        NumTabs & Number of tab characters \\
        NumSpaces & Number of space characters \\
        NumEmptyLines & Number of empty lines \\
        WhiteSpaceRatio & Ratio of whitespace characters \\
        NewLineBeforeOpenBrace & Newline character before open brace \\
        TabsLeadLines & Number of lines starting with a tab character \\
        \bottomrule
    \end{tabular}
\end{table}

\subsubsection{Syntactic Features}
Syntactic features are derived from the structure of the Code as represented by its Abstract Syntax Tree (AST) \cite{ast}. These features capture the deeper syntactic patterns and constructs used by the programmer. The following syntactic features were extracted:

\textit{Maximum Depth of AST Node:} Measures the maximum depth of any node in the AST.

\textit{Java Keywords:} Counts the occurrences of each Java keyword within the Code. Each Java code can contain one or more of these keywords, varying through different code snippets based on the Java keywords the programmer used during programming.

As we have many syntactic features, please refer to the project's documentation on its repository \cite{pordanesh} for more information on extracted syntactic features.

\subsection{Evaluator Model Training}
Evaluator model training is a fundamental aspect of this research project, providing the mechanism to assess the effectiveness of the style-altered Code generated by the large language models (LLMs). This section outlines the rationale behind the model selection, the detailed training process, and the validation of the model to ensure its robustness and reliability.

\subsubsection{Model Selection and Rationale}
The selection of the Random Forest classifier was driven by several compelling factors. Firstly, the effectiveness of Random Forest \cite{randomforest} in similar applications has been well-documented. Notably, the foundational study by Caliskan-Islam et al. \cite{caliskan1} utilized Random Forest with significant success in de-anonymizing programmers based on code stylometry. This provided a strong precedent for our choice. Random Forest classifiers are also known for their robustness to overfitting and provide valuable insights into feature importance.

\subsubsection{Training Process and Parameters}
The training process for the evaluator model was carefully designed to ensure robustness and accuracy. The pipeline began with \textit{feature extraction}, where a comprehensive set of 109 stylometric features was extracted from the primary dataset, categorized into lexical, layout, and syntactic types. To handle missing data, a simple \textit{imputation} strategy was employed, replacing any missing values with zeros. Subsequently, the dataset was prepared for \textit{data splitting} into training and testing subsets using an 80-20 ratio. Stratified sampling was used to maintain the distribution of the target variable (\texttt{repo\_id}) across both subsets. The \textit{model training} itself involved configuring a Random Forest classifier with 100 estimators (decision trees) and a random state of 42 to ensure reproducibility. Finally, to further assess robustness, a \textit{5-fold cross-validation} was performed on the training data.

\subsubsection{Model Validation}
Model validation is crucial to ensure that the evaluator model performs well not only on the training data but also on unseen data. The validation process included several metrics to comprehensively assess performance. The \textit{overall test accuracy} on the holdout set was a strong 95.18\%. The \textit{cross-validation accuracy scores} further demonstrated the model's consistency, yielding scores of 0.9447, 0.9504, 0.9508, 0.9511, and 0.9389, for a mean cross-validation accuracy of 0.95 with a standard deviation of 0.01. A \textit{detailed classification report}, presented in Table~\ref{tab:performance_metrics}, provided additional insights, where key metrics such as \textbf{Precision} (accuracy of positive predictions), \textbf{Recall} (ability to capture all relevant instances), and the \textbf{F1-score} (the balanced mean of precision and recall) confirmed the model's effectiveness.

\begin{table}[!t]
    \centering
    \caption{Evaluator Model Performance Metrics}
    \label{tab:performance_metrics}
    
    \begin{tabular}{lcccc}
        \toprule
        \textbf{Metric} & \textbf{Precision} & \textbf{Recall} & \textbf{F1-Score} & \textbf{Support} \\
        \midrule
        Accuracy & & & 0.95 & 13005 \\
        Macro Average & 0.92 & 0.89 & 0.90 & 13005 \\
        Weighted Average & 0.95 & 0.95 & 0.95 & 13005 \\
        \bottomrule
    \end{tabular}
\end{table}

\subsection{Prompt Engineering}
\begin{figure}[H]
    \centering
    \includegraphics[width=1\columnwidth]{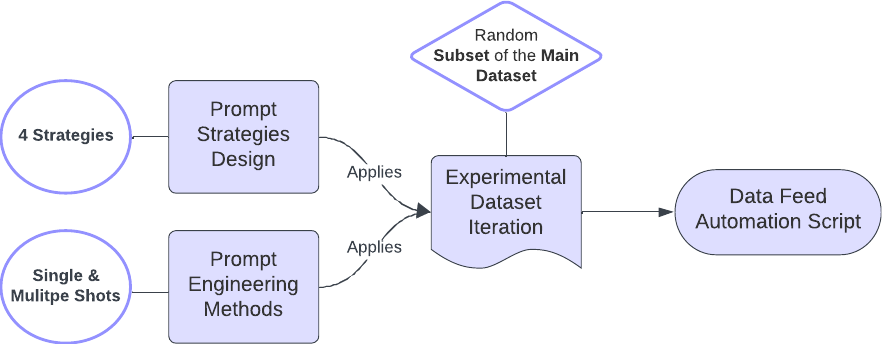}
    \caption{Overview of the prompt engineering and strategies designing.}
    \label{fig:experiment_design3}
\end{figure}

Prompt engineering is a critical component of our project, designed to interact with LLMs in a structured and effective manner to alter the stylistic elements of Code while preserving its functionality. The process is illustrated in Figure~\ref{fig:experiment_design3}.

\subsection{Prompt Engineering and Experimental Setup}
To evaluate the LLMs' ability to alter code stylometry, we developed a structured prompting methodology. Four distinct strategies were designed, with each strategy being implemented via two methods: a direct **single-shot** request and a more guided, conversational **multi-shot** request. This approach allowed for a comprehensive analysis of how prompt design influences stylistic transformation.

The following sections detail each strategy and the specific prompts used.

\subsubsection{Strategy 1: Just Change Style}
This baseline strategy tested the LLM's intrinsic ability to alter code style without specific guidance. The model was simply asked to rewrite the code to remove any identifying author signatures while preserving functionality.

\begin{promptbox}
\textbf{Single-Shot Prompt:}\\
\texttt{Please modify the following code snippet to change its style... The goal is to alter stylistic elements like variable naming, syntax structure, and code comments without affecting the functionality...}

\vspace{1em} 

\textbf{Multi-Shot Steps:}\\
\begin{enumerate}
    \item \texttt{Analyze the following Code to identify its key functional components and outputs.}
    \item \texttt{List alternative ways to implement the same functionality...}
    \item \texttt{Select one alternative... and rewrite the original Code...}
\end{enumerate}
\end{promptbox}

\subsubsection{Strategy 2: Specific Style Emulation}
This strategy assessed the LLM's capacity to adopt a specified persona. The prompt guided the model to mimic the coding style of a well-known programmer or to adhere to a regional coding convention.

\begin{promptbox}
\textbf{Single-Shot Prompt:}\\
\texttt{The following code snippet should be transformed to mimic the coding style of [famous programmer/region-specific coding style]. Adjust variable names, syntax, and code structuring to reflect this style...}

\vspace{1em}

\textbf{Multi-Shot Steps:}\\
\begin{enumerate}
    \item \texttt{Identify key stylistic traits of [famous programmer/region]...}
    \item \texttt{Outline how the current Code can be adapted to incorporate these traits...}
    \item \texttt{Rewrite the Code according to the outlined adaptations...}
\end{enumerate}
\end{promptbox}

\subsubsection{Strategy 3: Style with an Example}
This strategy employed a one-shot learning approach, where the LLM was provided with an example code snippet exhibiting the target style. This tested the model's ability to extract and apply stylistic patterns from a given sample.

\begin{promptbox}
\textbf{Single-Shot Prompt:}\\
\texttt{Here is a code snippet alongside an example... Modify the first Code to match the style of the example, including variable naming conventions, syntax, and structuring...}

\vspace{1em}

\textbf{Multi-Shot Steps:}\\
\begin{enumerate}
    \item \texttt{Compare the provided code snippet with the example to identify key stylistic differences.}
    \item \texttt{Create a plan to adapt the snippet's style to match the example...}
    \item \texttt{Implement the plan, rewriting the original code snippet...}
\end{enumerate}
\end{promptbox}

\subsubsection{Strategy 4: Cross-Language Transformation}
The final strategy aimed to introduce stylistic changes through a translation process. The code was translated from its original language (Java) to an intermediate language (C++) and then translated back, forcing the model to make new stylistic choices during the reverse translation.

\begin{promptbox}
\textbf{Single-Shot Prompt:}\\
\texttt{Translate the following Java code into C++... After translation, revert the C++ code to Java, applying any stylistic changes encountered during translation...}

\vspace{1em}

\textbf{Multi-Shot Steps:}\\
\begin{enumerate}
    \item \texttt{Convert the given Java code into C++...}
    \item \texttt{Review the C++ code to identify any stylistic... changes...}
    \item \texttt{Translate the C++ code back into Java, incorporating the identified changes...}
\end{enumerate}
\end{promptbox}

\subsection{LLM Communication and Execution Pipeline}
\begin{figure}[H]
    \centering
    \includegraphics[width=1\columnwidth]{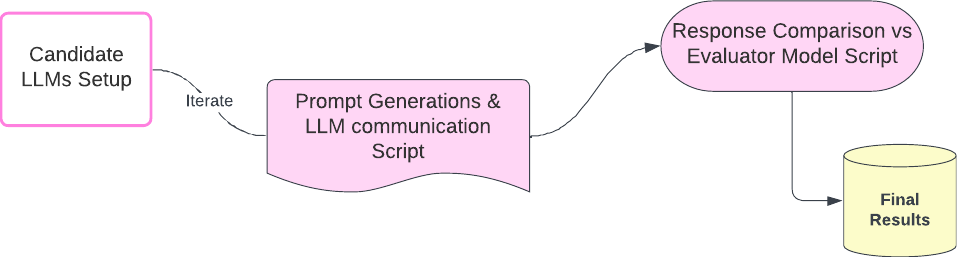}
    \caption{Overview of the LLM communication pipeline and result gathering.}
    \label{fig:experiment_design4}
\end{figure}

The experimental pipeline (Figure~\ref{fig:experiment_design4}) was fully automated. For each of the 10 code samples, scripts generated and dispatched the prompts for all strategies to both GPT-3.5 and GPT-4. The resulting 160 generated code snippets (10 samples $\times$ 4 strategies $\times$ 2 methods $\times$ 2 LLMs) were then programmatically evaluated against our trained stylometry classifier to measure the effectiveness of the style obfuscation.

\subsection{Evaluation Metrics}
The evaluation metrics are crucial for assessing the performance of the LLM-generated code in terms of both stylometric alteration and functionality preservation.

\subsubsection{Accuracy of Code Stylometry Detection}
This metric measures how effectively the LLMs alter the code style so the evaluator model cannot correctly identify the original \texttt{repo\_id}. The evaluation pipeline involved three main steps. First, the same set of 109 stylometric features was extracted from each LLM-generated code sample. Next, any missing values in the feature set were handled through imputation, replacing them with zeros to ensure a complete vector. Finally, the feature vector was sent to the evaluator model to predict the \texttt{repo\_id}. The ultimate success of a style transfer was determined by a binary \textbf{Pass/Fail} metric, based on whether the predicted \texttt{repo\_id} matched the original.

\subsubsection{Functionality Preservation of Altered Code}
This second metric assesses whether the LLM-generated code maintains its original functionality after the stylistic alterations. The verification was conducted through a rigorous compilation and testing process. For each generated sample, the original code component in its repository was replaced with the LLM-altered version. The entire repository was then recompiled to check for integration errors, and if successful, the project's full suite of unit tests was executed to confirm operational integrity.

The outcome for each sample was categorized into one of three states: successful compilation with all unit tests passing (\texttt{p/p}), failure during compilation (\texttt{fc}), or successful compilation followed by one or more failing unit tests (\texttt{p/f}). This functionality preservation test was performed on a subset of the generated samples to ensure a balance between stylistic obfuscation and functionality preservation.

\section{Results}
This section presents a detailed comparative analysis of the results obtained from the Large Language Models (LLMs), GPT-3.5 \cite{gpt-3.5} and GPT-4 \cite{gpt4}, and their effectiveness in altering code stylometry to evade detection by the evaluator model. The main results table summarizes the performance of different strategies and methods used in this study.

\subsection{Main Table of Raw Results}
The main results table captures the performance of various prompt strategies applied to the GPT-3.5 and GPT-4 models. Each entry indicates whether the evaluator model successfully identified the \texttt{repo\_id} after the LLM altered the code style.

\begin{table*}[!t]
    \centering
    \small 
    \caption{The Main Table of Raw Evasion Results (1 = Detection, 0 = Evasion)}
    \label{tab:main_results}
    \setlength{\tabcolsep}{4pt} 
    \begin{tabular}{lcccccccccc}
        \toprule
        \textbf{Model, Strategy \& Method} & \textbf{r1} & \textbf{r2} & \textbf{r3} & \textbf{r4} & \textbf{r5} & \textbf{r6} & \textbf{r7} & \textbf{r8} & \textbf{r9} & \textbf{r10} \\
        \midrule
        \texttt{gpt-3.5-turbo\_s1s} & 1 & 1 & 1 & 1 & 1 & 1 & 1 & 1 & 1 & 1 \\
        \texttt{gpt-3.5-turbo\_s2s} & 1 & 0 & 1 & 1 & 1 & 1 & 1 & 1 & 1 & 0 \\
        \texttt{gpt-3.5-turbo\_s3s} & 1 & 1 & 1 & 1 & 0 & 0 & 1 & 1 & 1 & 1 \\
        \texttt{gpt-3.5-turbo\_s4s} & 1 & 1 & 1 & 0 & 0 & 1 & 1 & 1 & 1 & 1 \\
        \texttt{gpt-4-turbo\_s1s}   & 1 & 1 & 1 & 1 & 1 & 1 & 1 & 0 & 1 & 1 \\
        \texttt{gpt-4-turbo\_s2s}   & 0 & 1 & 1 & 1 & 0 & 1 & 1 & 1 & 1 & 0 \\
        \texttt{gpt-4-turbo\_s3s}   & 1 & 1 & 1 & 1 & 1 & 1 & 1 & 1 & 1 & 1 \\
        \texttt{gpt-4-turbo\_s4s}   & 1 & 1 & 0 & 0 & 0 & 1 & 1 & 1 & 1 & 1 \\
        \addlinespace
        \texttt{gpt-3.5-turbo\_s1m} & 0 & 0 & 1 & 1 & 0 & 0 & 1 & 0 & 1 & 0 \\
        \texttt{gpt-3.5-turbo\_s2m} & 1 & 1 & 1 & 1 & 0 & 1 & 1 & 1 & 1 & 1 \\
        \texttt{gpt-3.5-turbo\_s3m} & 1 & 0 & 1 & 1 & 0 & 1 & 1 & 1 & 1 & 1 \\
        \texttt{gpt-3.5-turbo\_s4m} & 1 & 0 & 1 & 0 & 0 & 1 & 1 & 1 & 1 & 1 \\
        \texttt{gpt-4-turbo\_s1m}   & 1 & 1 & 1 & 1 & 0 & 1 & 1 & 0 & 1 & 0 \\
        \texttt{gpt-4-turbo\_s2m}   & 1 & 0 & 0 & 1 & 0 & 1 & 1 & 1 & 0 & 0 \\
        \texttt{gpt-4-turbo\_s3m}   & 1 & 1 & 0 & 1 & 0 & 1 & 1 & 1 & 1 & 1 \\
        \texttt{gpt-4-turbo\_s4m}   & 0 & 1 & 1 & 0 & 0 & 0 & 1 & 1 & 1 & 0 \\
        \bottomrule
    \end{tabular}
\end{table*}

\subsubsection{Understanding the Results}
To facilitate comprehension of the main results table, the format is explained as follows.
Each row is labeled with the format \texttt{[model]\_[strategy][method]}, where for instance, \texttt{gpt-3.5-turbo\_s4s} indicates the GPT-3.5 model using the fourth strategy (S4) with a single-shot method ('s').
The columns, labeled \texttt{r1} to \texttt{r10}, represent the performance on each of the 10 code samples.
Finally, the cell entries signify the outcome: a \textbf{1} indicates the evaluator model correctly predicted the \texttt{repo\_id}, meaning the LLM failed to sufficiently alter the style. Conversely, a \textbf{0} indicates the evaluator failed to make a prediction, meaning the LLM successfully evaded detection.

\subsection{Competitive Analysis}
The aggregated results are presented here to facilitate a comprehensive understanding of the effectiveness of the different LLMs and prompt strategies. The following table captures the percentage of successful style evasions (represented by '0's in the main table) for each combination. A higher percentage indicates a greater success rate for the LLM.

\begin{table}[!t]
    \centering
    \caption{Percentage of Successful Evasions (out of 10 samples)}
    \label{tab:analysis_table}
    \begin{tabular}{lcccccccc}
        \toprule
        \textbf{Method} & \multicolumn{4}{c}{\textbf{GPT-3.5}} & \multicolumn{4}{c}{\textbf{GPT-4}} \\
        \cmidrule(lr){2-5} \cmidrule(lr){6-9}
        & \textbf{S1} & \textbf{S2} & \textbf{S3} & \textbf{S4} & \textbf{S1} & \textbf{S2} & \textbf{S3} & \textbf{S4} \\
        \midrule
        Single & 0\% & 20\% & 20\% & 20\% & 10\% & 40\% & 0\% & 30\% \\
        Multi  & 60\% & 10\% & 20\% & 40\% & 30\% & 50\% & 20\% & 50\% \\
        \bottomrule
    \end{tabular}
\end{table}

\subsubsection{Prompt Strategy Effectiveness}
This section evaluates the effectiveness of different prompting methods. For the single-shot method, GPT-4 with strategy S2 achieved the highest success rate at 40\%, while GPT-3.5 with strategy S1 showed the lowest at 0\%, with an average success rate across all single-shot experiments of 14\%. These results, detailed in Table~\ref{tab:single_shot_results}, indicate that GPT-4 models were generally more effective in single-shot methods.

\begin{table}[!t]
    \centering
    \caption{Single-Shot Method Performance Results}
    \label{tab:single_shot_results}
    \begin{tabular}{lcc}
        \toprule
        \textbf{Strategy} & \textbf{GPT-3.5-turbo} & \textbf{GPT-4-turbo} \\
        \midrule
        S1 & 0\% & 10\% \\
        S2 & 20\% & 40\% \\
        S3 & 20\% & 0\% \\
        S4 & 20\% & 30\% \\
        \bottomrule
    \end{tabular}
\end{table}

The multi-shot method, which provides more detailed instructions, was generally more effective, with an average success rate of 28\%. As shown in Table~\ref{tab:multi_shot_results}, the highest success rate of 60\% was achieved by GPT-3.5 with strategy S1, while its lowest performance was 10\% with strategy S2. This indicates that detailed, step-by-step guidance can significantly enhance performance, particularly for GPT-3.5.

\begin{table}[!t]
    \centering
    \caption{Multi-Shot Method Performance Results}
    \label{tab:multi_shot_results}
    \begin{tabular}{lcc}
        \toprule
        \textbf{Strategy} & \textbf{GPT-3.5-turbo} & \textbf{GPT-4-turbo} \\
        \midrule
        S1 & 60\% & 30\% \\
        S2 & 10\% & 50\% \\
        S3 & 20\% & 20\% \\
        S4 & 40\% & 50\% \\
        \bottomrule
    \end{tabular}
\end{table}

\subsubsection{Prompt Scenario Effectiveness}
This section analyzes the effectiveness of the four core prompt scenarios. The Cross-Language Processing strategy (S4) was the most effective overall, with an average success rate of 35\%, as detailed in Table~\ref{tab:success_rate_by_scenario}. The "Change Style to a Specific Style" (S2) strategy was also relatively effective at 30\%. The "Just Ask to Change the Style" (S1) strategy demonstrated mixed effectiveness (25\%), with its success being highly dependent on the use of multi-shot prompts. The least effective strategy was "Change Style with an Example" (S3), with a success rate of 15\%, indicating that the models had difficulty generalizing from a single provided example.

\begin{table}[!t]
    \centering
    \caption{Average Success Rate by Prompt Scenario}
    \label{tab:success_rate_by_scenario}
    \begin{tabular}{lc}
        \toprule
        \textbf{Scenario} & \textbf{Average Success Rate} \\
        \midrule
        Just Ask to Change the Style (S1) & 25\% \\
        Change Style to a Specific Style (S2) & 30\% \\
        Change Style with an Example (S3) & 15\% \\
        Cross-Language Processing (S4) & 35\% \\
        \bottomrule
    \end{tabular}
\end{table}

\subsubsection{LLM Effectiveness}
This section compares the overall effectiveness of the two LLMs. The GPT-4-turbo model slightly outperformed its predecessor, achieving an average success rate of 29\% compared to 24\% for GPT-3.5-turbo (Table~\ref{tab:model_success_rate}). GPT-3.5 demonstrated more consistent performance in single-shot methods but achieved higher success with the detailed guidance of multi-shot methods, particularly in the S1 strategy. In contrast, GPT-4 showed higher variability in single-shot methods but notable success in the S2 and S4 multi-shot strategies, emphasizing its capability to adapt to detailed, structured prompts.

\begin{table}[!t]
    \centering
    \caption{Average Success Rate by Model}
    \label{tab:model_success_rate}
    \begin{tabular}{lc}
        \toprule
        \textbf{Model} & \textbf{Average Success Rate} \\
        \midrule
        GPT-3.5-turbo & 24\% \\
        GPT-4-turbo & 29\% \\
        \bottomrule
    \end{tabular}
\end{table}

\subsection{Functionality Preservation Check Results}
This check aimed to evaluate whether LLM-modified code retained its functionality. This involved compiling the modified code within its original repository and running unit tests.

\subsubsection{Compilation and Unit Test Results}
We were able to compile and test samples from 5 of the 10 repositories, as dependency issues prevented testing on some older repositories. This resulted in a total of 80 code snippets being tested (5 repositories $\times$ 16 modifications each). The outcomes, summarized in Table~\ref{tab:integrity_success_rate}, show that 58\% of modified snippets compiled and passed all tests. However, a significant portion (41\%) failed to compile, often due to LLM-induced changes to class names, function names, or data types. A single sample (1\%) compiled but failed its unit tests.

\begin{table}[!t]
    \centering
    \caption{Functionality Preservation Check Results (N=80)}
    \label{tab:integrity_success_rate}
    \begin{tabular}{lcc}
        \toprule
        \textbf{Test Outcome} & \textbf{Count} & \textbf{Percentage} \\
        \midrule
        Successfully Compiled \& Passed Tests & 46 & 58\% \\
        Failed to Compile & 33 & 41\% \\
        Compiled but Failed Unit Tests & 1 & 1\% \\
        \bottomrule
    \end{tabular}
\end{table}

\subsubsection{Analysis of Successfully Compiled and Tested Code}
Further analysis was conducted on the 46 fully functional code snippets to determine their evasion effectiveness. As detailed in Table~\ref{tab:llm_effectiveness_final}, 8 of these 46 samples (17\%) also successfully evaded the stylometry classifier. This means that 10\% of the total set of tested snippets (8 out of 80) achieved the dual goal of this study: they both retained full functionality and successfully concealed their original authorship style.

\begin{table}[!t]
    \centering
    \caption{Evasion Success for Functionally Correct Code}
    \label{tab:llm_effectiveness_final}
    \begin{tabular}{lc}
        \toprule
        \textbf{Metric} & \textbf{Value} \\
        \midrule
        Functionally Correct Samples & 46 \\
        Of those, Evasion was Successful & 8 (17\%) \\
        \hline 
        \textbf{Overall Success Rate (Functional \& Evasive)} & \textbf{10\%} \\
        \textbf{(8 out of 80 total samples)} & \\
        \bottomrule
    \end{tabular}
\end{table}

\section{Discussion}

\subsection{Interpretation of Results}
\subsubsection{LLM Perspective}
The results from our experiments provide a detailed view of the performance and capabilities of GPT-3.5 \cite{gpt-3.5} and GPT-4 \cite{gpt4} in altering code stylometry. When analyzing the overall effectiveness of these models, it becomes apparent that the newer GPT-4 model generally outperforms GPT-3.5, especially in specific prompt strategies.

GPT-4 demonstrated a notable improvement in single-shot strategies, with a success rate peaking at 40\% for the strategy that changes style to a specific programmer or geographical style (S2). This suggests that GPT-4's more advanced capabilities enable it to better understand and apply complex stylistic changes even with minimal guidance. Conversely, GPT-3.5 struggled with single-shot methods, particularly with the simple "Just Ask" strategy (S1), which resulted in a 0\% success rate. This implies that GPT-3.5 may require more structured and detailed guidance to achieve effective stylistic transformations.

In multi-shot methods, both models showed improved performance. The highest success rate for GPT-3.5 was 60\% using the S1 strategy in a multi-shot approach, highlighting the importance of incremental and detailed instructions in helping the model achieve the desired stylistic changes. GPT-4 also showed strong performance with multi-shot methods, achieving up to 50\% success in specific strategies. This reinforces the model's ability to apply detailed, step-by-step guidance to alter code styles effectively.

\subsubsection{Prompt Engineering Perspective}
The effectiveness of the different prompt engineering strategies provides further insights into the LLMs' capabilities and limitations.

The performance of single-shot methods, which involve providing a single prompt, varied significantly. The results indicate that these methods are generally less effective, especially for GPT-3.5. The exception was GPT-4, which managed a 40\% success rate with the S2 strategy, suggesting that while single-shot prompts can be effective, they may require a more advanced model capable of interpreting complex stylistic changes from a single command.

In contrast, multi-shot methods, which broke down the task into multiple steps, proved more effective overall. This approach provides the LLM with incremental guidance, making it easier for the model to understand and apply the desired stylistic changes. The success of GPT-3.5 in the multi-shot approach (peaking at 60\%) underscores the importance of detailed instructions. Similarly, GPT-4's consistent performance across multi-shot strategies highlights its ability to efficiently handle more complex and structured tasks.

An analysis of the individual prompt strategies reveals further nuances:

\paragraph{Just Ask to Change the Style (S1)} This straightforward approach showed mixed effectiveness. The significant improvement in success rates for multi-shot prompts, particularly for GPT-3.5, indicates that detailed guidance is crucial for this strategy's effectiveness.

\paragraph{Change Style to a Specific Style (S2)} This strategy was relatively effective, especially with GPT-4 in both single and multi-shot methods. The structured nature of this strategy, which involves mimicking specific styles, likely provides clear and concrete guidance that the LLM can follow more easily.

\paragraph{Change Style with an Example (S3)} Consistent but lower success rates were observed for this strategy across all methods. This suggests that while providing an example can help, it may not be as effective as more direct and detailed guidance.

\paragraph{Cross-Language Processing (S4)} This strategy proved to be the most effective overall, with higher success rates, especially in multi-shot methods. This strategy's detailed, step-by-step nature likely aids the LLM in making the necessary stylistic adjustments more accurately.

\subsection{Limitations}
Despite the significant findings of this study, several limitations must be acknowledged to provide context for the results and to guide future research. These limitations are primarily related to dataset constraints, computational resources, prompt engineering challenges, and issues encountered during the code functionality preservation checks.

\subsubsection{Dataset Constraints}
One of the primary challenges was the difficulty in identifying a suitable dataset. A key constraint was the dataset's \textit{size and freshness}; finding a Java dataset with unit tests large enough for comprehensive results yet recent enough to reflect current coding practices proved challenging. Furthermore, while we aimed for \textit{diversity} in coding styles, the dataset's reliance on open-source projects may have introduced biases, as it might not fully represent the vast array of programming styles found in proprietary or less-documented codebases.

\subsubsection{Computational Resources}
The training and evaluation of the models used in this study required substantial computational resources. The evaluator model training, in particular, necessitated \textit{high-performance computing needs} due to the large dataset and numerous features. The Random Forest algorithm, while effective, is computationally intensive. Even with access to high-performance computing, \textit{resource allocation} was a constant challenge, limiting the ability to run multiple iterations or explore additional model configurations that could have provided more profound insights.

\subsubsection{Prompt Engineering Challenges}
Prompt engineering, a critical aspect of this study, faced notable challenges, primarily due to \textit{budget constraints}. The cost associated with API communications for LLMs like GPT-3.5 and GPT-4 imposed significant limitations, as each prompt incurred costs, particularly those with large tokens. This financial constraint led to \textit{limited exploration} of the extensive range of potential prompt engineering techniques, which may have otherwise provided deeper insights into the LLMs' performance.

\subsubsection{Code Functionality Preservation Check Issues}
Ensuring that the modified code maintained its functionality was a critical component of the study, yet several issues arose. The absence of an automated system necessitated \textit{manual compilation and testing} of each repository, which significantly increased the workload. This manual process inherently introduced the \textit{potential for human error}, which could affect the consistency of the functionality checks. Finally, this manual approach created \textit{scalability issues}, highlighting the need for automated solutions to efficiently handle larger datasets in future work.

\section{Future Work}
The findings of this study lay the groundwork for several avenues of future research. Addressing the limitations and exploring new methodologies will enhance the robustness and applicability of the results. The key areas for future work are outlined below.

\subsection{Expanded Datasets}
One of the primary limitations of the current study was the size and diversity of the dataset. Future work should focus on acquiring a larger, more diver

se, and up-to-date Java dataset with unit tests. Expanding the dataset will provide a broader range of coding styles and practices, which is crucial for training more effective evaluator models. Furthermore, a greater number of data samples must be tested against the LLMs by applying a broader array of prompt strategies and methods. This will yield more robust results than testing only 10 data samples.

\subsection{Improved Functionality Preservation Pipeline}
A significant limitation of the current study was the manual compilation and testing of modified code samples. Future work should focus on developing an automated pipeline for this process to streamline functionality checks and reduce the manual workload. Implementing an automated system will ensure consistency and reliability in the functionality preservation checks while also significantly reducing the time and effort required. This automation is key to improving the scalability of the experiments, allowing for more extensive testing across a larger number of repositories and code samples.

\subsection{Broader Exploration of Prompt Strategies}
The budget for API communication limited the range of prompt strategies that could be tested. Future work should focus on implementing and testing a wider variety of tailored and well-engineered prompt strategies. Developing more specific and nuanced prompts could enhance the effectiveness of code style modification, helping to achieve more subtle and sophisticated stylistic changes. Exploring a broader variety of strategies and scenarios will provide deeper insights into the capabilities and limitations of LLMs in altering code stylometry and will help identify the most effective approaches for different coding styles.

\subsection{Alternative Machine Learning Algorithms for Evaluator Model}
While the Random Forest algorithm used in this study was effective, exploring alternative machine learning algorithms could further improve the accuracy and robustness of the evaluator model. Investigating other techniques, such as Long Short-Term Memory (LSTM) \cite{lstm} networks, can provide different strengths in detecting code stylometry. LSTMs, in particular, are well-suited for sequential data and could capture patterns in code that other algorithms might miss.

\subsection{Inclusion of Other Enterprise LLMs}
Exploring and utilizing other advanced enterprise-level Large Language Models (LLMs) will provide a broader perspective on the capabilities and limitations of current AI technologies in this domain. Future work should include investigating models such as Gemini[], Claude[], LlaMa3[], and Phi[] to allow for a comprehensive comparison of their performance. Conducting a comparative analysis of these diverse LLMs will help to identify the best-performing models and the specific contexts in which they excel, providing valuable insights for future research and practical applications.

\section{Conclusion}
This study explored the effectiveness of GPT-3.5 and GPT-4 in altering code stylometry while preserving functionality. The results demonstrated that GPT-4 generally outperforms GPT-3.5, especially with specific and structured prompt strategies. Multi-shot methods proved more effective than single-shot methods, highlighting the importance of detailed guidance. The study also revealed significant challenges related to dataset constraints, computational resources, prompt engineering, and functionality preservation. Addressing these limitations and exploring new methodologies will enhance the robustness and applicability of future research in this area. The findings provide critical insights for developing more robust authorship attribution techniques and advancing the field of code stylometry in the face of evolving AI capabilities.

\printbibliography


\appendix

\section{Methods2Test Dataset JSON Structure}
\label{app:json_structure}

A.1 Methods2Test Dataset JSON Structure.

\begin{center}
    \includegraphics[width=0.9\columnwidth]{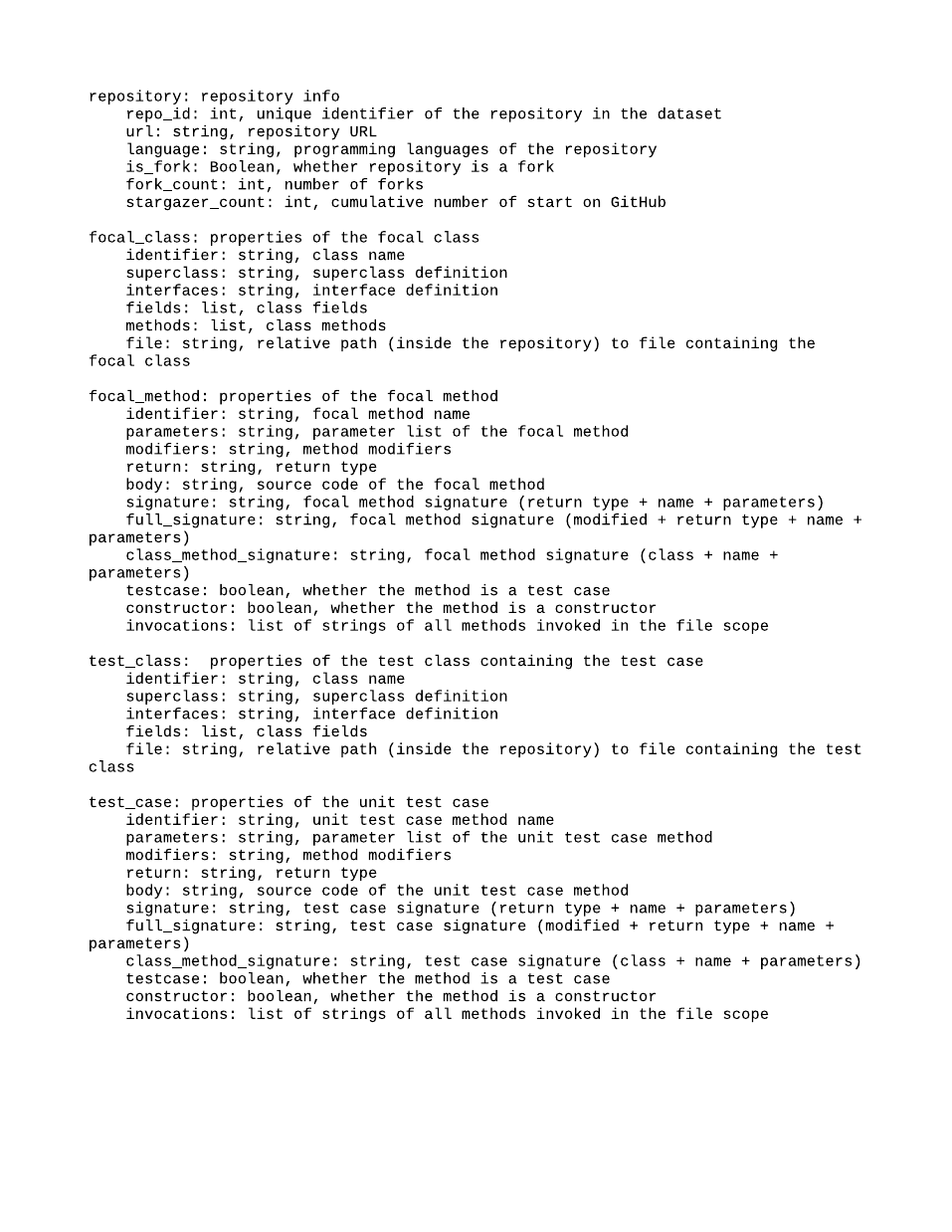}
    
    \captionof{figure}{Methods2Test Dataset JSON Structure.}
    \label{fig:JSON_structure}
\end{center}

\end{document}